\documentstyle[12pt,epsfig]{article}
\def\g{\gamma}

\def\bsg{b \rightarrow s \, \g}

\newcommand{\Bdsll}{B^0_{d,s} \to l^+ l^-}

\newcommand{\Bsll}{B^0_s \to l^+ l^-}
\newcommand{\Bdll}{B^0_d \to l^+ l^-}

\newcommand{\Bsmm}{B^0_s \to \mu^+ \mu^-}
\newcommand{\Bdmm}{B^0_d \to \mu^+ \mu^-}

\newcommand{\Bdtt}{B^0_d \to \tau^+ \tau^-}

\def\beq{\begin{equation}}
\def\eeq{\end{equation}}
\def\bea{\begin{eqnarray}}
\def\eea{\end{eqnarray}}
\def\nnb{\nonumber}
\def\ed{\end{document}}

\newcommand{\gsim}{\lower.7ex\hbox{$\;\stackrel{\textstyle>}{\sim}\;$}}
\newcommand{\lsim}{\lower.7ex\hbox{$\;\stackrel{\textstyle<}{\sim}\;$}}

\begin{document}
\begin{flushright}
\baselineskip=12pt
IC/2001/174\\
hep-ph/0201121
\end{flushright}
\vskip 0.2cm

\begin{center}
{\LARGE\bf
$(g-2)_{\mu}$ and CP Asymmetries in $B^0_{d,s} \to l^+ l^-$ and $b \to
s \gamma$ in SUSY models}

\vspace{0.2cm}
Chao-Shang Huang$^a$,~ LIAO Wei$^b$\\ 
\vspace{0.2cm}
        $^a$ ITP, Academia Sinica, P.O. Box 2735, 100080 Beijing, China \\
        $^b$ ICTP, Strada Costiera 11, 34014 Trieste, Italy
\end{center}
\begin{abstract}
We show that with a good fit to the muon $g-2$ constraint,
the CP asymmetry can be as large as $25\%$ ($15\%$) for
$B^0_d \to \tau^+ \tau^-$ ($B^0_d \to \mu^+ \mu^-$) in
SUGRA models with nonuniversal gaugino masses and MSSM.
If tau events indentified at B factories the CP asymmetry
in $B^0_d \to  \tau^+\tau^-$ can be a powerful probe on 
physics beyond SM. An interesting case is that new physics 
gives a prediction on the branching ratio which is still in 
the uncertain region of the SM prediction. The CP asymmetry 
for $B^0_d \to l^+ l^-$ in this case can still reach $20\%$. 
So it is powerful to shed light on physics beyond SM while 
the CP asymmetry of $b \to s \gamma$ in this case can only 
reach $2\%$ at most which is too small to draw a definite 
conclusion on new physics effects at B factories.
\end{abstract}

The FCNC process $\Bdsll$ has been shown in recent years
to be a powerful process to shed light on new physics beyond SM
~\cite{dhh,hy,bk,ALOT,ddn,hky}
especially for SUSY models which may enhance the decay amplitude by
$\tan^3\beta$\cite{hy,bk}, provided that $\tan\beta$ is large (say, 
$ \gsim 20$). It became more interesting recently after the Brookhaven
National Laboratory(BNL) reported the $2.6 \sigma$ excess of the
muon anomalous magnetic moment $a_\mu=(g-2)_\mu/2$ over its SM value:
$\Delta a_{\mu}\equiv a_\mu^{exp}-a_\mu^{SM}=(43\pm16)10^{-10}$\cite{BNL}. 
SUSY models with large $\tan\beta$ (say \gsim 10) are favored by this excess.
It was shown for mSUGRA in \cite{ddn}
that with a good fit to $(g-2)_{\mu}$ the branching ratio(Br) of $\Bdsll$
can be enhanced by 100 times and within good reach at Tevatron
Run II. Over the last two months the theoretical prediction of $a_{\mu}$
in the standard model (SM) has undergone a significant revision due to the
change in sign of the light by light hadronic correction to $a_{\mu}$, which
leads to 
\bea
\Delta a_{\mu}=26(16)\times 10^{-10}\label{g-2}
\eea
corresponding to a $1.6\sigma$
deviation from SM~\cite{sign}. In this letter we will examine the CP violating
effects of the process in the light of this new result of $\Delta a_{\mu}$.

 While there is no direct CP violation for this process, there might be
CP violation induced by mixing of $B^0$ and $\bar{B}^0$ in the process 
\begin{equation}
B^0\rightarrow\bar{B}^0\rightarrow f~~~~~~~vs.~~~~~~~~\bar{B}^0
\rightarrow B^0\rightarrow \bar{f}. \nnb\\
\end{equation}
One can define the CP violating observable as
\bea
 A_{CP}= 
\frac{\int^\infty_0 dt ~\sum_{i=1,2}\Gamma(B^0_{phys}(t) \rightarrow f_i)
-\int^\infty_0 dt ~\sum_{i=1,2}\Gamma(\bar{B}^0_{phys}(t) \rightarrow \bar{f}_i)}
{\int^\infty_0 dt ~\sum_{i=1,2}\Gamma(B^0_{phys}(t) \rightarrow f_i)
+\int^\infty_0 dt ~\sum_{i=1,2}\Gamma(\bar{B}^0_{phys}(t) \rightarrow \bar{f}_i)},
\label{cp}
\eea
where $f_{1,2}=l^+_{L,R}l^-_{L,R}$ with $l_{L(R)}$ being the 
helicity eignstate of eigenvalue $-1 (+1)$,
$\bar{f}$ is the CP conjugated 
state of $f$, i.e. $\bar{f}_{1,2}=l^+_{R,L}l^-_{R,L}$. 
In the approximation $|{q \over p}|=1$ it is
\bea
A_{CP} &=& - \frac{2 Im(\xi) X_q}
{(1+ |\xi|^2)(1+X_q^2)}, ~~~q=d,s,\label{app}
\eea
where $X_q= \frac{\Delta m_q}{\Gamma}$($q=d,s$ for $B^0_d$ and $B^0_s$
respectively), $p$ and $q$ are parameters diagonalising the B meson mass matrix
: $|B^0_{L,H}>= p |B^0>\pm q |\bar{B}^0>(|p|^2+|q|^2=1$), and
$\xi$ is
\bea
\xi &=&
 \frac{C_{Q1}\sqrt{1-4\hat{m}_l^2}
+ (C_{Q2}+2 \hat{m}_l C_{10})}{C^*_{Q1}\sqrt{1-4\hat{m}_l^2}
- (C^*_{Q2}+2 \hat{m}_l C^*_{10})}\label{rate}
\eea
with $\hat{m}_l = m_l/m_{B^0}$.
In Eq. (\ref{rate}) $C_{Qi}$'s are Wilson coefficients accounting for neutral
Higgs contributions, and $C_{10}$ is the Wilson coefficient for the
axial vector operator (for details, see refs.\cite{dhh,hy,ALOT}\footnote{Note that
the explicit expressions of the Wilson coefficients $C_{Q_i}$'s are the same in SUSY
models with and without CP violating phases. In the SUSY models with CP violating
phases the coefficients become complex since the new CP violating phases enter into
squark and chargino mass matrices~\cite{hl1}.}). In deriving Eq. (\ref{app}) we have
used $
\frac{q}{p}= - \frac{M^*_{12}}{|M_{12}|}= - \frac{\lambda^*_t}{\lambda_t}$
which is the result for $B^0$-$\bar{B}^0$ mixing in the SM. This is a 
good approximation in the MSSM if one limits himself to the regions 
with large $\tan\beta$ ( say, larger than 10 but smaller than 60 ), not too
light charged Higgs boson ( say, larger than 250 Gev ), and heavy sparticles,  
and in the scenarios of the minimal flavor violation (MFV)
\footnote{MFV means the models in
which the CKM matrix remains the unique source of flavor violation. The models have further be
subdivided into the so-called MFV models and the
generalized MFV models (GMFV) in ref.\cite{bcrs}}
 without new CP violating phases there is also no correction 
to the result~\cite{bcrs,ir}.
In MFV models with new CP violating phases, e.g., in the 
SUGRA with nonuniversal
gaugino masses which we consider in the latter, a rough estimate 
gives that the correction to
the SM value of $q/p$ is below $20 \%$ in the parameter space we 
used in the latter. The detailed
analysis of the effects on $B-{\bar B}$ mixing of new CP violating 
phases in such models 
will be given in the forthcoming paper~\cite{hlw}.
It is easy to see from Eq. (\ref{app}) that the
kinetic upper bound of $A_{CP}$ is $X_q/(1+X_q^2)$ which is about
$48\%$ for $B^0_d$ and $6.3\%$ for $B^0_s$. 
The SM predictions for these observables are of order $10^{-3}$($10^{-4}$) 
for $B^0_d$($B^0_s$) if corrections to $|\frac{q}{p}|=1$ included\cite{hl}. 
In our approximation they are zero in SM. We shall concentrate on the 
decay modes $\Bdll$ in the latter
and it is straightforward to apply to $\Bsll$. 

Before presenting numerical results, we can make an estimate on
the magnitude of the CP asymmetry in the process.
For the case of SUSY contributions dominated(e.g.
Br enhanced by 100 times larger) the main
contributions to $C_{Qi}$'s are from the FCNC self-energy type diagram
with neutral Higgs bosons coupled to the external bottom quark\cite{hy,ALOT}.
So one may get $m^2_{H0} C_{Q1} \approx -m^2_{A0} C_{Q2}$ (the light neutral
Higgs contribution is not important in general by observing that the light
neutral Higgs should decouple to resemble the SM Higgs if the SUSY spectrum
is a little of bit high or $\tan\beta$ is large).
Consequently for $l=e, \mu$ we have $|\xi| \approx |(m^2_{H0}-m^2_{A0})|
/(m^2_{H0}+m^2_{A0})$. This is a small number in most regions of
the parameter space of MSSM,
since in the case of large $m_{A0}$ ( $m_{A0}^2\gg m_Z^2$ )
$m_{H0}$ and $m_{A0}$ are aligned
with just a few percents discrepancies at most.
This estimate is polluted in two ways. One is that penguin and box
diagrams give a deviation from the relation $m^2_{H0} C_{Q1}= -m^2_{A0} C_{Q2}$
which should be less than $5\%$ for $\tan\beta$ larger than 20. Another one
is given by the appearance of $2 {\hat m}_\mu C_{10}$ in Eq. (\ref{rate}) 
which is about $-0.18$ and can give about $10\%$ deviation. Interferences
among all of these can give the CP asymmetry for $\Bdmm$ of $15\%$ if the 
phase of $\xi$ is large ( $\sim \pi/2$ ). Obviously this estimation
is not valid for $l=\tau$ since the factor $\sqrt{1-4\hat{m}_\tau^2}
\approx 0.74$ is now important. CP asymmetry can then be increased
by about $10\%$ due to this factor.

For muon $g-2$, it has been shown that in most of the parameter
space in SUSY models the chargino exchange dominates the
contribution\cite{ce} which is proportional to $\sum_j \tan\beta
/m_{\chi_j^\pm} Re(V^*_{j1}U^*_{j2})$ (where V, U are the matrices
diagonalising the chargino mass matrix and the loop function is not shown) in the
case of large $\tan\beta$. 
SUGRA models with nonuniversal gaugino masses have been reexamined
\cite{icn} in the light of the muon $g-2$ constraint. 
It was shown that $60-90\%$ of the regions of the parameter space
are excluded by it. They are mainly regions with
low $\tan\beta$, as expected. In the following we will consider 
scenarios with large $\tan\beta$($> 10$) and discuss the BNL constraint
on the SUGRA model with nonuniversal gaugino masses.
The values of CP violating phases in the model are obtained by satisfying the
constraints from the eletric dipole moments (EDMs)
of the electron and of the neutron based on the cancellation
mechanism\cite{IN,hl1}. In the SUGRA model with nonuniversal 
gaugino masses~\cite{hl1,nonu}, compared with mSUGRA
\footnote{In mSUGRA there in general
are only four real parameters (the universal scalar and gaugino masses 
$M_0$ and $M_{1/2}$, the trilinear term $A_0$ and $\tan\beta$) and two 
CP violating phases (the phase of the Higgsino mass parameter $\mu$ 
($\phi_\mu$) and the phase of $A_0$).}, there are two more real parameters 
(say, $|M_1|$ and $|M_3|$, where $M_1$ and $M_3$ are gauge masses 
corresponding to $U(1)$ and $SU(3)$ respectively), and two
more independent phases arising from complex gaugino masses, which 
make the cancellations among various SUSY contributions to EDMs easier 
than in mSUGRA. Therefore relatively large values of the phase of $\mu$ 
are allowed. It was shown in \cite{hl1} 
that in the SUGRA model $\phi_\mu$ is bounded to be about $0.5$
at most (as in \cite{hl1} we work in the convention that $B \mu$ and
$M_2$ are real and the remaining phases of $A_0$, $M_1$, $M_3$ and
$\mu$ are then considered physical) while the phases of gaugino 
masses can be of order one
and give interesting implications on CP violation of B decays.
\footnote{Essentially the effects of the phases of gaugino masses
on $C_{Q_i}$'s (and consequently CP violation
in B decays) come from the running of $A_t$ since $A_t$ 
at the EW scale mainly depends on $M_3$ at the high energy (GUT) scale
through RGE effects. For details, see ref.~\cite{hl1,hlyz}.}
Accordingly, it is expected that in the SUGRA model the correlation 
between $Br(\Bdsll)$ and $a^{SUSY}_\mu$ which was shown for CP conserving 
mSUGRA in \cite{ddn} will be
altered just a little by the presence of $\phi_\mu$ in the chargino
mass matrix and the qualitative properties should be the same,
that is, for a reasonable value of $a^{SUSY}_\mu$ we can get a quite
large enhancement of $Br(B^0_{d,s}\rightarrow l^+l^-)$ over its SM value. 
Our numerical calculations confirmed the expectation.

As well known, inclusive $\bsg$ is also sensitive to $\tan\beta$ in SUSY
models and the CP asymmetry in $\bsg$ can be large (say $10\%$) in models 
beyond SM~\cite{kn,BK}, which is well under the probe at B factories.
So while studying the CP violation of $\Bdsll$ we should
also consider $\bsg$ and we would like to know what implications on $\Bdsll$
we can get if the CP asymmetry in $\bsg$ is measured to a good precision.
The CP asymmetry in $\bsg$ is
\bea
  && a_{\rm CP}^{b\to s\gamma}(\delta)
   = \frac{\Gamma(\bar B\to X_s\gamma)-\Gamma(B\to X_{\bar s}\gamma)}
            {\Gamma(\bar B\to X_s\gamma)+\Gamma(B\to X_{\bar s}\gamma)}
    \Bigg|_{E_\gamma>(1-\delta) E_\gamma^{\rm max}} \nonumber\\
  &&=\frac{\alpha_s(m_b)}{|C_7|^2}\,\Bigg\{
    \bigg[\frac{40}{81}- \frac{8z}{9}
    \Big[ v(z) + b(z,\delta) \Big]\bigg]
    \mbox{Im}(C_2 C_7^*) 
   \mbox{}- \frac 49\,\mbox{Im}(C_8 C_7^*)
    + \frac{8z}{27}\,b(z,\delta)\,\mbox{Im}(C_2 C_8^*)
    \Bigg\} ~~~
\label{ACP}
\eea
where $\sqrt{z}=m_c/m_b=0.29$, $v(z)$ and $b(z,\delta)$ can be found 
in \cite{kn}. 
Among all of the terms the important are interferences
between $C_7$ and $C_2$ and between $C_8$ and $C_7$ (in the following we 
denote this CP asymmetry as {\bf $a_{CP}$} to avoid the confusion with that 
of $\Bdsll$).  It is the imaginary part of $C_7$ that plays the main role. Because $Im(C_7)$
for which SUSY contributions are responsible linearly depends on $\tan\beta$
in the large $\tan\beta$ case, we expect that if $a_{CP}$ is observed
to be large (note that in this case the $2\sigma$ Br bound, 
$2.0 \times 10^{-4} < Br(\bsg) <4.5 \times 10^{-4}$\cite{CLEO1}, 
can be satisfied through the cancellation between the SM and SUSY 
contributions), $a^{SUSY}_\mu$ must also be large. 
$A_{CP}$, in this case,
should also be significant by observing that SUSY contributions to $C_{Qi}$'s
share the vertex similar to that contributing to $C_7$.
However, large enough $A_{CP}$ does not necessarily imply large enough 
$a_{CP}$. This is because $C_{Qi}$'s (depending on $\tan^3\beta$)
can easily dominate the process $\Bdsll$ while $\tan\beta$ is not
large enough to make $a_{CP}$ significant.
Until now the result on $a_{CP}$ is still preliminary:
$-0.27 < A(\bsg) <0.10$ at $90\%$ CL\cite{CLEO2}.

\begin{figure}[t]
\centerline{\psfig{figure=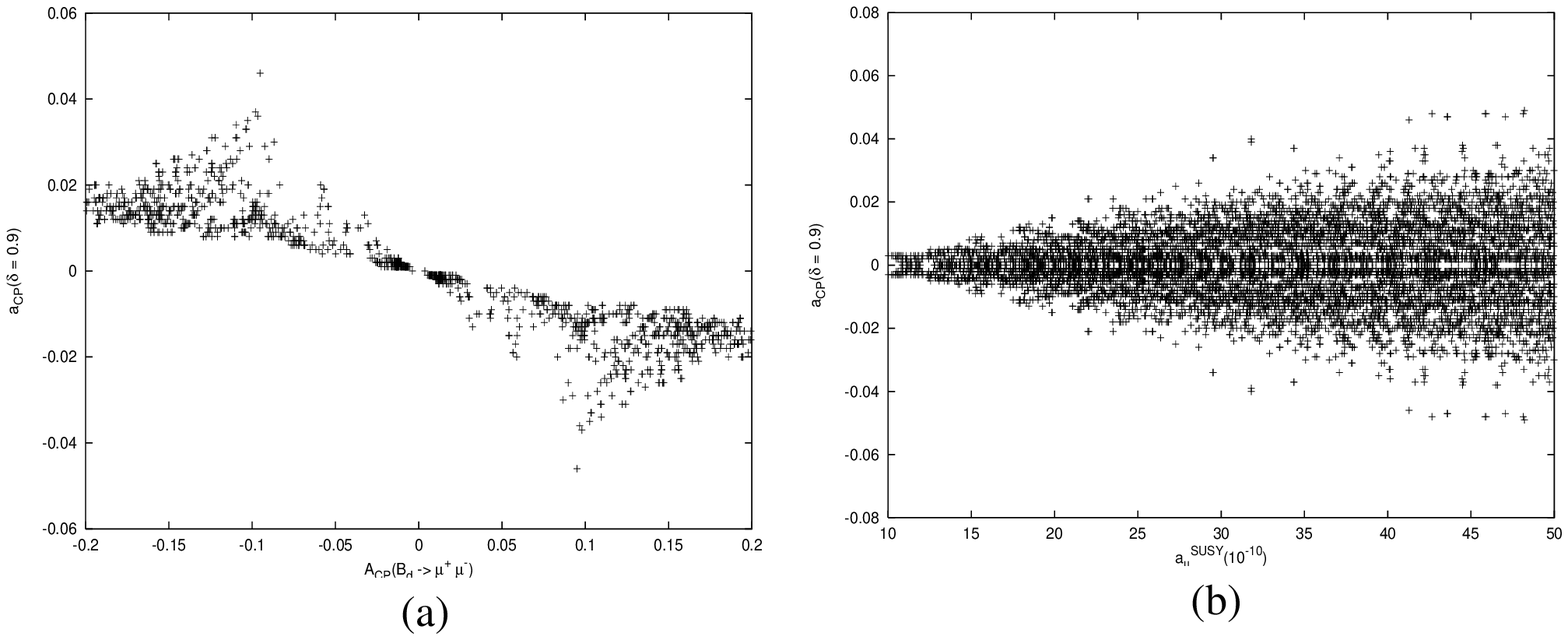,height=8cm,width=18cm}}
\caption{\small (a) $a_{CP}(\bsg)$ versus $A_{CP}(\Bdmm)$ in the parameter
space $150 <|M_1|<350 $,$150 <M_2 <300$, $150 <|M_3|<500 $,
$|A_0|=300 $, $M_0=250 $ and $10<\tan\beta<50$;(b) $a_{CP}(\bsg)$
versus $a_\mu^{SUSY}$.
  }
\label{fig1}
\end{figure}

Our results are shown in Figs. 1 and 2.
In Fig.\ref{fig1}(a) we give a plot for the correlation between the CP asymmetries
of $\Bdmm$ and $\bsg$ (for $\delta=0.9$) with points scanned in the region
as shown in the figure caption (in which the long-lived chargino mass bound is satisfied
 and the neutral LSP constraint included). Dimensionful parameters 
are with unit GeV. Besides the $\bsg$ Br bound, experimental
constraints also considered include the muon $g-2$ bound (the result (\ref{g-2})
with a 1$\sigma$ error corridor),
$10 \times 10^{-10}< \Delta a_\mu(=a_{\mu}^{SUSY})<42 \times 10^{-10}$,
the Br bound of $\Bsmm$, $Br(\Bsmm)<2.0 \times 10^{-6}$($90\%$ CL)\cite{PDG}
(the bound from $\Bdmm$ is weaker and not relevant). For the lightest Higgs boson
mass $m_{h^0}$ we set $m_{h^0}\geq 112$ GeV\footnote{In the autumn
of year 2000 a $2 \sigma$ excess of Higgs boson signal was reported
at energy of about $115$GeV.
Now the result with complete data has appeared\cite{sopczak}.
It is shown that there is a narrow region in which the $m_{h0}$
and $m_{A0}$ are allowed to both reach $89$ GeV. Beyond this region
$m_{h0}$ and $m_{A0}$ are bounded from below by  $112-113$GeV. To stress
the importance of our study we therefore leave that narrow region
and make a conservative study
by assuming that the Higgs boson mass should
not be less than $112$ GeV.}. In calculating the Br and CP asymmetries
we have included the higher loop corrections due to the large bottom Yukawa
coupling in the large $\tan\beta$ case~\cite{cgnw}.
The plot is shown for $Br(\Bdmm)$ enhanced by at least 
5 times larger than that of SM, i.e., $\gsim 7.5 \times 10^{-10}$.  One may see
that two asymmetries do not exhibit a strong correlation. 
Fig.\ref{fig1}(b) is devoted to the correlation
of $a_{CP}$ and $a^{SUSY}_\mu$ without the requirement 
$Br(\Bdmm)\gsim 7.5 \times 10^{-10}$. One may find that $5\%$ $a_{CP}$ 
requires more than a $2\sigma$ deviation of $a_\mu$\footnote{We also calculate
the correlation of $a_{\mu}$ with $A_{CP}$. The result is that for a $1\sigma$ deviation 
of $a_{\mu}$, $A_{CP}$ can reach $10\%$.}. 
Back to Fig.\ref{fig1}(a), one may find that
for $a_{CP}$ of about $2\%$ $A_{CP}$ can still reach $10\%$.
This means that one may use $A_{CP}(\Bdmm)$ to determine
CP phases if $a_{CP}$ is just about $2\%$ which is hard to give an
answer on new physics because of the about $1\%$ SM prediction\cite{kn}. It's
also clear that for $a_{CP}$ just above $2\%$ $A_{CP}$ would be inevitably
large ($\gsim 10\%$). This is exactly what we expected. Unfortunately, the Br
of this process is too small. Even in Tevatron Run IIb experiments
with 20 fb$^{-1}$ integrated luminosity collected, only several tens events
can be got for Br enhanced by 100 times. One should wait LHCb
experiments to get an answer on it. We have also calculated CP violation in
$\Bsmm$ in the same region as that shown in the caption of Fig. 1 and the result is
that the CP asymmetry can reach about $1.5\%$.

Problems will be overcome if tau events are identified since 
$B \to \tau^+\tau^-$ has the Br over 160 times larger than that of
$B \to \mu^+ \mu^-$.
For a quite large part of points shown in Fig.\ref{fig1}(a),
corresponding $A_{CP}(\Bdtt)$ can reach $25\%$. If 900 events collected, 
for example, for which with the 20 fb$^{-1}$ integrated luminosity the Br only needs to be 
enhanced by about 30 times of the SM prediction, then even
all the errors including those of b-tagging, tau-tagging and the statistical
altogether are of about $8\%$  which is quite large, one may still determine the
CP asymmetry, $A_{CP}$, to a $3 \sigma$ level. Since tau decays before it 
reaches detector, the helicity eigenstates can be measured according to the
final lepton energy. The CP asymmetries defined referring to definite
helicity, i.e. $A_{CP}^i$'s ( i=1,2 )\cite{hl}, can then be useful.
Because $|\xi|$ is in general a number less than one, $A^1_{CP}(\Bdtt)$
is small because of  $X_d \approx 0.74$, while the magnitude of
$A^2_{CP}(\Bdtt)$ is larger than 
$A_{CP}$ and  its kinematic maximum is $63\%$\cite{hl}.
Basically $A_{CP}$ and $A^2_{CP}$ can give independent implications
on the CP violating phases. However
at the Tevatron only for the case of $Br(\Bdtt)$ 
enhanced by two orders of magnitude over its SM value it is possible to probe
$A^2_{CP}$ to a good precision. The reason is that 
Br($B^0_d \to \tau^+_R \tau^-_R$) is proportional to
$|C_{Q1}\sqrt{1-4\hat{m}_\tau^2}+ (C_{Q2}+2 \hat{m}_\tau C_{10})|^2$
which is suppressed by an order of magnitude compared with the 
total Br because of the cancellation between $C_{Q1}$ and $C_{Q2}$,
as we noted. 

\begin{figure}[t]
\centerline{\psfig{figure=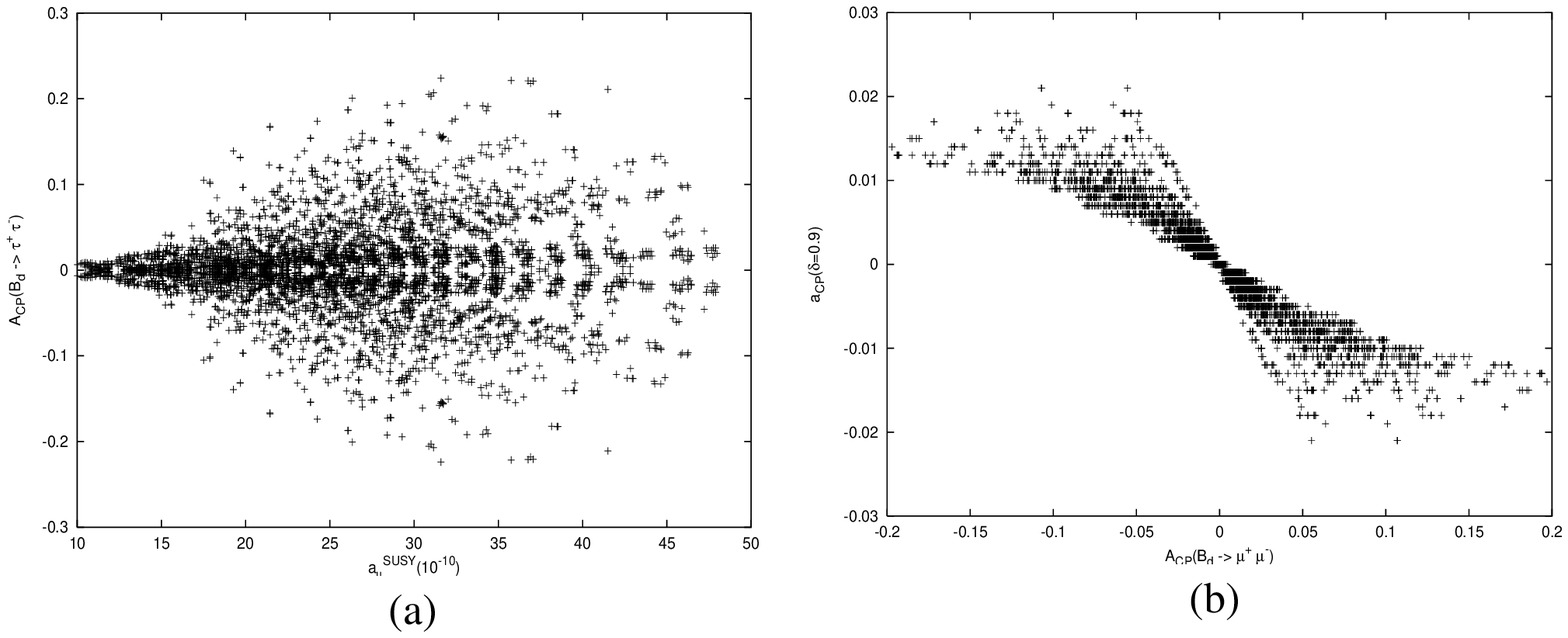,height=8cm,width=18cm}}
\caption{\small (a) $a_\mu^{SUSY}$ versus $A_{CP}(\Bdtt)$ with 
$Br(\Bdtt)<5.0 \times 10^{-8}$; (b) $a_{CP}(\bsg)$
versus $A_{CP}(\Bdmm)$.
  }
\label{fig2}
\end{figure}

In the above analysis we assume that the Br of the process is 
increased a lot compared with that in the SM. This possibility
can be probed by searching for $\Bsmm$ at Tevatron Run II. In fact there
are large regions of the parameter space, e.g., for $\tan\beta$ of $10-20$,
in which the Br can not be increased too much. In particular, there is
a possibility that the Br is increased at most by just about $60\%$ 
compared with that of SM 
which is in the uncertain region of the SM prediction. For example,
for $\Bdtt$ if enhanced by $60\%$ then the Br is 
($4.9\pm1.6)\times 10^{-8}$. It still has overlap with the SM prediction:
($3.1\pm1.0)\times 10^{-8}$. An especially interesting
question then is whether one can still learn something
about new physics from its CP violating effects if the Br of the process should be 
confirmed to be in the uncertain region of the SM prediction. To answer this question,
we plot the correlations of $A_{CP}(\Bdtt)$ with $a^{SUSY}_\mu$ in 
Fig.\ref{fig2}(a), setting the calculated center value of the 
Br to satisfy  $Br(\Bdtt) < 5.0 \times 10^{-8}$,
scanning the parameter space as shown in Fig.\ref{fig1}.
The answer is obviously yes.
CP asymmetry can reach $20\%$ allowed by the muon $g-2$ constraint 
within $2\sigma$ deviations, 
 which could be probed at the LHCb. We would like to note that
$A_{CP}(\Bdmm)$ in this case may not be reduced by about $10\%$ compared
with $A_{CP}(\Bdtt)$ and can 
also reach $20\%$. Fig.\ref{fig2}(b) is devoted to the correlation between
$A_{CP}(\Bdmm)$ and $a_{CP}(\bsg)$.  We may observe
that $a_{CP}$ can reach only $2\%$ while $A_{CP}$ can be as large as $20\%$.
In this case $Br(B^0_d \to \tau^+_R \tau^-_R)$ is generally of the same order
of the total Br($\Bdtt$). Consequently $A^2_{CP}$ and $A_{CP}$
would exhibit a strong correlation. 

We have done analyses in the SUGRA model with non-universal gaugino
masses. For MSSM which has many more CP violating phases,  
 the qualitative features of our results are also valid at least in
the scenario of minimal flavor violation (MFV). 
As shown in ref.~\cite{in}, the $a_{\mu}^{SUSY}$ dependence on phases from both
the chargino and neutralino exchanges consists only of three combinations
which depend on the phases of $\mu, A_{\mu}$ and gaugino masses. Compared with
the SUGRA model we used above, there is one more phase, the phase of $A_{\mu}$,
which enters $a_{\mu}^{SUSY}$ through the neutralino exchange.
 Because the regions of the parameter space we 
considered in the letter, as noted before, are the regions where the chargino exchange
contribution dominates, it makes no significant effects on our result. Although the presence 
of phases of $A_{u,d,e}$ makes larger $\phi_\mu$ possible which may reduce $a_\mu^{SUSY}$, 
 it is still hard for $\phi_\mu$ to reach one in the large 
$\tan\beta$ case so that the effects of the phases of $A_{u,d,e}$ on 
$a_\mu^{SUSY}$ are small and consequently negligible.
It is obvious that these more phases have no significant effects on 
the CP asymmetries in $\bsg$ and $\Bdsll$ since the main contributions 
to the processes $\bsg$ and $\Bdsll$
come from the third generation quarks and their superpartners\footnote{If relaxing to
the scenario of non MFV, e.g., there are non-diagonal trilinear terms, then the
non-diagonal trilinear terms will make significant effects on the CP asymmetry in $\bsg$~\cite
{nmfv}.}. 
Therefore, we conclude that qualitatively our results should be valid in quite large
regions of the parameter space in MSSM.

We stress that CP asymmetry of $\Bdsll$ is theoretically very clean
because the decay constant, which is the only source of hadronic 
uncertainty for the process, is cancelled out for the asymmetry. 
The above analysis have shown that CP violation of $\Bdtt$ is highly 
interesting for finding new physics and could be observed at Run II of 
the Tevatron if tau events can be identified.  Based on the expectation 
we encourage experimentalists to do a good job on tau-tagging.

In the analysis so far we didn't 
include the effects from the mixing of CP-odd and CP-even
neutral Higgs bosons in SUSY models\cite{cepw}.
This effect is expected to increase the mass difference
between $A^0$ and $H^0$ in the presence of CP violating phases.
Roughly $(m^2_{H0}-m^2_{A0})/(m^2_{H0}+m^2_{A0})$ can be increased up
to $10\%$ for $m_{H\pm}$ of $200-300$ GeV, which implies larger
CP asymmetry in the $\Bdsll$ process.

In conclusion, we have shown the correlation of the CP asymmetry in 
$\Bdsll$ with the muon anomalous magnetic moment $a_{\mu}$
and the CP asymmetry in $b \to s \gamma$ in SUSY models.
Besides the branching ratio, the CP asymmetry $A_{CP}$ of the
$\Bdsll$ (l=$\mu, \tau$) process is also highly interesting
to shed light on physics beyond SM. Compared with the branching ratio, 
theoretically CP asymmetry of $\Bdsll$  is of more advantages because 
the common uncertain decay constant cancels out in the definitions
of the CP violating observables. 
CP asymmetry of $\Bdll$ can be as large as several tens
percents while the constraints from the muon anomalous magnetic moment 
and $b\rightarrow s \gamma$ are satisfied. If large CP violating effects
were to be found out in the $\bsg$ process, CP asymmetry of
$\Bdll$ would be inevitably large. On the other hand for smaller than
$2\%$ CP asymmetry of $\bsg$, $\Bdll$ can still have CP violating
effects of $10\%$($20\%$) for $l=\mu(\tau)$. If tau events identified
with $6\%$ tagging error, one can measure $A_{CP}(\Bdtt)$ to a $3\sigma$
level at Run II of Tevatron with Br($\Bdtt$) enhanced by a factor of about 30 
compared to that of SM. It is interesting when the nature prefers a 
scenario for which new physics only
increases the branching ratio a little so that it is still in
the uncertain region of the SM prediction. CP asymmetry of $\Bdll$
in this scenario can still reach $20\%$ allowed by the muon $g-2$ constraint 
within $2\sigma$ deviations. Therefore it can provide a powerful way to 
probe new physics beyond the SM, while CP violation of $\bsg$ can only be 
about $2\%$ at most which is not large enough to probe new physics since the
prediction for CP violation of $\bsg$ in SM is of order $1\%$ in magnitude\cite{kn}. 

The work is supported in part by the National Natural Science
Foundation of China.

\end{document}